\documentclass{revtex4}
\usepackage{amssymb}
\usepackage{amsmath} 
\usepackage{graphicx}

\begin{document}

\title{\bf  Dirac particle in a spherical scalar potential well}
\author{R. Layeghnejad$^1$}   \email{r.layeghnejad@mail.yu.ac.ir}
\author{M. Zare$^1$}
\author{R. Moazzemi$^2$}
\affiliation{
$^1$Department of Physics, Yasouj University, Yasouj 75919-353, Iran\\
$^2$Department of Physics, The University of Qom,  Ghadir Blv., Qom 371614-611, Iran}

\date{\today}
\begin{abstract}
In this paper we investigate a solution of the Dirac equation for a spin-$\frac{1}2$ particle in a scalar potential well with full spherical symmetry. The energy eigenvalues for the quark particle in $s_{_{1\hspace{-.4mm}{\diagup_{\hspace{-.8mm}2}}}}$  states (with
$\kappa=-1$) and  $p_{_{1\hspace{-.4mm}{\diagup_{\hspace{-.8mm}2}}}}$ states (with
$\kappa=1$) are calculated. We also study the continuous Dirac wave function for a quark in such a potential, which is not necessarily infinite. Our results, at infinite limit, are in good agreement with the MIT bag model. We make some remarks about the sharpness value of the wave function on the wall. This model, for finite values of potential, also could serve as an effective model for the nucleus where $U(r)$ is the effective single particle potential.
\end{abstract}

\maketitle

\section{Introduction}

In 1964  Gell-Mann and Zweig independently proposed a quark model in which all known hadrons were described as bound states of only three fundamental particles \cite{gell,zw}. Each of these spin-1/2 particles, which  Gell-Mann named quarks, has their corresponding antiparticles. However, a precise mechanism of the bound states and quark confinement has still not been established. Gell-Mann's model requires that the valid quark states should be colorless and so a free quark is not allowed. Since a free single quark has not yet been observed, the model assumes that quarks appear as bound states. The baryons are bound states of three quarks (or antiquarks) and the mesons are made up of one quark and one antiquark. As a model, to describe this behavior, one can think of a hadron as a cavity in which the quark wave function is confined, and the cavity is surrounded by the QCD vacuum \cite{[10]}. Therefore in a simple model quarks confined in a sphere with radius $R$ and the Dirac equation sould be solved within this sphere.

From this point of view, the solution of the Dirac equation in a three-dimensional scalar potential, apart from its interesting theoretical aspects, provide useful tools for studying the properties of elementary particles. For example, the MIT bag model \cite{[5a],[5b],[5c]} and its charily invariant versions, such as the chiral bag model \cite{[6],[7a],[7b]} and the cloudy bag model \cite{[8a],[8b],[8c],[8d]}, are some models for describing the physics of the nucleon and other baryons. A bag is a region of space in which quarks and gluons are confined, i.e. they are forced by an external constant pressure B, which can be fitted using experimentally determined hadron masses, to move only inside the bag. Historically, Chodos {\it{et. al.}} have considered the MIT bag model through the Dirac equation (except the bag pressure B) \cite{[5a]}. They solved bag equations for the massless Dirac fields in three space dimensions. Their solutions are for the special case of static spherical boundary. They also computed charge radius and found it to be 1.0 fm. Degrand {\it{et. al.}}, in the other case of this model, calculated the masses and the static parameters of the light hadrons \cite{[5b]}. In the cloudy bag model a baryon is treated as a three-quark bag that is surrounded by a cloud of pions. Thomas {\it{et. al.}} investigated the static properties of the nucleon within this model \cite{[8a]}. They found the bag radius to be about 0.8 fm  by a fit to pion-nucleon scattering in the (3,3)-resonance region. The chiral bag model for the nucleon is a hybrid of quark and meson degrees of freedom, interpolating the two limits of the skyrme model at $R\to 0$  and the MIT bag model at $R\to\infty$ \cite{[6],[54],[55]}. In the skyrme model mesons acts as gauge particles so that baryons would interact with each other by the exchange of mesons \cite{d}. Skyrmions are the solutions of the field equations. These solutions are solitons and no longer plain wave. One may interpret these skyrmions as coherent states of baryons and excited baryons \cite{a}; however, the physical interpretation is still not completely resolved. With this model, it is also possible to calculate nucleon masses and other  particle properties \cite{b}. Both the MIT bag and skyrme model are useful to calculate masses and other properties of hadrons. In Ref. \cite{[6]}, Hosaka and Toki investigated the static properties of the nucleon such as masses and magnetic moments as a function of $R$, in both the original chiral bag model and models with vector mesons.
The MIT bag model introduces many free parameters for energy corrections that could be helpful in understanding the physical processes  inside the nucleus.

It is a curious and complex situation to solve the relativistic quantum mechanics problems in a finite potential in comparison with the equivalent problem in nonrelativistic quantum mechanics.
In the Dirac equation, the wave function is continuous, its first derivative is discontinuous and the second derivative has a very large jump, whereas, in the Schr\"{o}dinger equation the wave function and its first derivative are continuous, but the second derivative has a certain jump related to the potential jump. The solutions of the Dirac equation in a $\delta$ potential exist in the literature, see for example \cite{[1],[2],[3],[4]}.
However, for a finite spherical potential well, as far as we are aware of, there is no solution in the literature.

In this paper, we consider the solution of the Dirac equation in a spherically symmetric scalar potential well, which is not necessarily infinite. The origin of this scalar potential could be a strong force that binds quarks together in clusters to make more familiar subatomic particles, such as protons and neutrons. It also holds together the atomic nucleus and underlies interactions between all particles containing quarks \cite{fritz,nobel}. Although this model is not compatible with the quark confinement at finite potential, it could serve a dual purpose. First, one could observe the evolution of the wave functions as $U_0\to\infty$, where one could recover the MIT bag model results. Second, for finite values of $U_0$ this could serve as an effective model for the nucleus where $U(r)$ is the effective single particle potential emerging from the meson exchange of the nucleons. Here we obtain eigenvalue equations for the energies and numerically calculate the energy eigenvalues for the  $s_{_{1\hspace{-.4mm}{\diagup_{\hspace{-.8mm}2}}}}$ and  $p_{_{1\hspace{-.4mm}{\diagup_{\hspace{-.8mm}2}}}}$ states. These states are derived normally from the continuity of the wave function and imposing boundary conditions on the cavity \cite{[9]}, with different radii $R=0.8,1,1.18$ fm and the quark masses $m=0,1$ fm$^{-1}$. We then, compare the values of the energy levels in each of specific conditions and also with the energy eigenvalue obtained in the previous MIT bag model.
The relation between energy eigenvalues and the radius of cavity and the mass of the quark is considered.
We also obtain the Dirac wave function components for a quark particle and depict them in figures. As a result, when the strength of the potential is increased, the wave function components on the boundary of cavity fall down and it would have the sharper point, which are in good agreement with the MIT bag model . Finally, we obtain the magnitude of sharpness of the relativistic wave function component when crossing the wall.

The paper is organized as follows: Section \ref{sec2} is devoted to an introduction of the Dirac equation with a central scalar potential. In Sec \ref{sec3} we calculate the energy eigenvalues for a Dirac particle in a scalar potential with full spherical symmetry for the $s_{_{1\hspace{-.4mm}{\diagup_{\hspace{-.8mm}2}}}}$  and  $p_{_{1\hspace{-.4mm}{\diagup_{\hspace{-.8mm}2}}}}$ states. Then we  discuss about the  sharpness of the wave function components on the boundary. Finally in Sec \ref{sec4} we summarize our results.

\section{Solutions of the Dirac equation in a central potential (scalar coupling)}\label{sec2}

In the Dirac equation, the scalar potential $U(x)$ and the fourth component of a vector potential, $V_0(x)$ are accompanied by mass $m$ and energy $E$, respectively. Although $V_0(x)$ is not a vector potential, since it is the fourth component of a four vector, it is called a vector potential. The general form of the equation of motion for a spin-$1/2$ particle with these two potential is (in relativistic units, $\hbar=1$ and $c = 1$)
\begin{equation}\label{1}
    [\alpha .p + \mathop \beta \limits^{} (m + U(x)) + V_0 (x)]\psi (x,t) = i\frac{\partial }{{\partial t}}\psi (x,t),
\end{equation}
where $p=-i\nabla$ is the three-dimensional momentum operator. In the above equation $\alpha$ and $\beta$ are the $4\times4$ Dirac matrices which, in the usual representation, are given by
\begin{equation}\label{2}
    \alpha=\left(
             \begin{array}{cc}
               0 & \sigma_i \\
               \sigma_i & 0 \\
             \end{array}
           \right),\qquad \beta=\left(
                                  \begin{array}{cc}
                                    0 & I \\
                                    -I & 0 \\
                                  \end{array}
                                \right),
\end{equation}
where $I$ is the $2\times2$  unit matrix. The subscript $i$ can take the values of  1,2,3, and $\sigma_i$ are  the $2\times2$ Pauli matrices.
For a Dirac particle in a spherically symmetric potential field, the total angular momentum operator $J$, and the spin-orbit matrix operator $K =  - \beta \left( {\sigma .L + 1} \right)$, commute with the Dirac Hamiltonian. Here $L$ is the orbital angular momentum. The complete set of the conservative quantities with their eigenvalues can be written as follows:
\begin{equation}\label{3}
 \begin{array}{ll}
 H \to E \qquad
 &K \to  - \kappa  \\
 J^2  \to j\left( {j + 1} \right) \qquad
 &J_3  \to j_3,  \\
 \end{array}
\end{equation}
so that,
\begin{equation}\label{4}
\left[ {H,K} \right] = 0,\qquad\left[ {H,{J}} \right] = 0,\qquad
\left[ {J^2 ,J_3 } \right] = 0,\qquad\left[ {{J},K} \right] = 0,
\end{equation}
\begin{equation}\label{5}
\quad \mbox{and}\quad\kappa  =  \pm \left( {j + \frac{1}{2}} \right),\quad\mbox{for}\quad l=j\mp1/2.
\end{equation}
Therefor the quantum number $\kappa$ is a nonzero integer number. Given stationary solutions $\psi _{jj_3 }^\kappa  \left( {x,t} \right) = \psi _{jj_3 }^\kappa  \left( x \right)\,e^{ - iEt}$,
 we have
\begin{equation}\label{6}
\psi _{jj_3 }^\kappa  (x) = \left( \begin{array}{l}
 g_\kappa  (r)y_{jl}^{j_3 }  \\
 if_\kappa  (r)y_{jl' }^{j_3 }  \\
 \end{array} \right),
\end{equation}
where $g_\kappa (r)$ and $f_\kappa (r)$ are real square-integrable functions, and $y_{jl}^{j_3 }$
 and $y_{jl'}^{j_3 }$ can be written in terms of the spherical harmonic functions with the relevant Clebsch-Gordan coefficients. Then the two coupled equations for the radial parts of the Dirac equation with a given scalar potential turn out to be,
\begin{eqnarray}
\label{7} \frac{{df_\kappa  (r)}}{{dr}} + \frac{{1 - \kappa }}{r}f_\kappa  (r)& = &\left( {m + U(r) - E} \right)g_\kappa  (r) \\\label{7.1}
 \frac{{dg_\kappa  (r)}}{{dr}} + \frac{{1 + \kappa }}{r}g_\kappa  (r) &=& \left({m + U(r) + E} \right)f_\kappa  (r).
 \end{eqnarray}

 The solutions of Eqs. (\ref{7}) and (\ref{7.1}) for a scalar potential well, $U(r\leq R)=0$ and $U(r>R)=U_0$, are the spherical Bessel functions and modified spherical Bessel functions of the first kind, for the regions  $r\leq R$ and $r>R$, respectively. Therefore
for region I, $r<R$ with $U(r) = 0$, and we have
\begin{subequations}\label{8}
\begin{gather}\label{8a}
\mbox{for}\;\kappa<0:\qquad
\left\{ \begin{array}{l}
 g_\kappa  (r) = Nj_{|\kappa| - 1} (pr) \\
 f_\kappa  (r) = - N\frac{p}{{ {m + E}}}j_{|\kappa|} (pr), \\
 \end{array} \right.
\\\nonumber\\\label{8b}
\mbox{for}\;\kappa>0:\qquad
\left\{ \begin{array}{l}
 g_\kappa  (r) = N'j_\kappa  (pr) \\
 f_\kappa  (r) = N'\frac{p}{{{m + E} }}j_{\kappa  - 1} (pr), \\
 \end{array} \right.
 \end{gather}
\end{subequations}
and for region II,  $r>R$ with $U(r) = U_0$,
\begin{subequations}\label{10}
\begin{gather}
\label{10a}
\mbox{for}\;\kappa<0:\qquad
\left\{ \begin{array}{l}
 g_\kappa  (r) = M K_{|\kappa| - 1} (qr) \\
 f_\kappa  (r) =  - M \frac{q}{{m + U_0  + E}} K_{|\kappa|} (qr), \\
 \end{array} \right.
\\\nonumber\\\label{10b}
\mbox{for}\;\kappa>0:\qquad
\left\{ \begin{array}{l}
 g_\kappa  (r) = M' K_\kappa(qr) \\
 f_\kappa  (r) =  - M' \frac{{q}}{{m + U_0  + E}} K_{\kappa  - 1} (qr), \\
 \end{array} \right.
  \end{gather}
\end{subequations}
where $p=\sqrt {E^2  - m^2 }$ and $q=\sqrt {\left( {m + U_0 } \right)^2  - E^2 }$, and $N, N', M$, and $M'$ are the normalization factors  \cite{[9],[10],[11]}.

\section{A Dirac particle in a potential well with full spherical symmetry}\label{sec3}
In this section we compute the energy eigenvalues for a Dirac particle in a full spherically symmetric scalar potential.
Reported eigenvalues, for which the massless spin-$1/2$ field is confined to an infinite spherical potential well of radius $R=1$ fm, are listed in Table \ref{table1}. These values have been computed through an equation of motion and boundary conditions \cite{[5a]}.

\begin{table}[th]
\begin{tabular}{|c|c|c|c|c|}
  \hline
  state&  $1s_{_{1\hspace{-.4mm}{\diagup_{\hspace{-.8mm}2}}}}$&$2s_{_{1\hspace{-.4mm}{\diagup_{\hspace{-.8mm}2}}}}$&
  $1p_{_{1\hspace{-.4mm}{\diagup_{\hspace{-.8mm}2}}}}$ &$2p_{_{1\hspace{-.4mm}{\diagup_{\hspace{-.8mm}2}}}}$ \\\hline
  $E \mbox{ (fm}^{-1})$&2.04& 5.40 & 3.81 & 7.00 \\
  \hline
\end{tabular}\caption{Energy eigenvalues for a massless quark in an infinite potential well with $R=1$ fm.\label{table1}}
\end{table}
The solution of the Dirac equation for a particle with mass $m$, which moves in a spherically symmetric static cavity, is physically similar to a scalar potential well with the full spherical symmetry. To confine the wave function of the particle in a bag,  as the vector potential is zero $V_0=0$, the depth of the well should be infinite. This method, which is applicable as $U_0\to\infty$, has been developed by the MIT bag model \cite{[10]}.
However, here we consider the scalar potential which is not \emph{necessarily} infinite.
\begin{figure}[th]
\includegraphics[width=5.5cm]{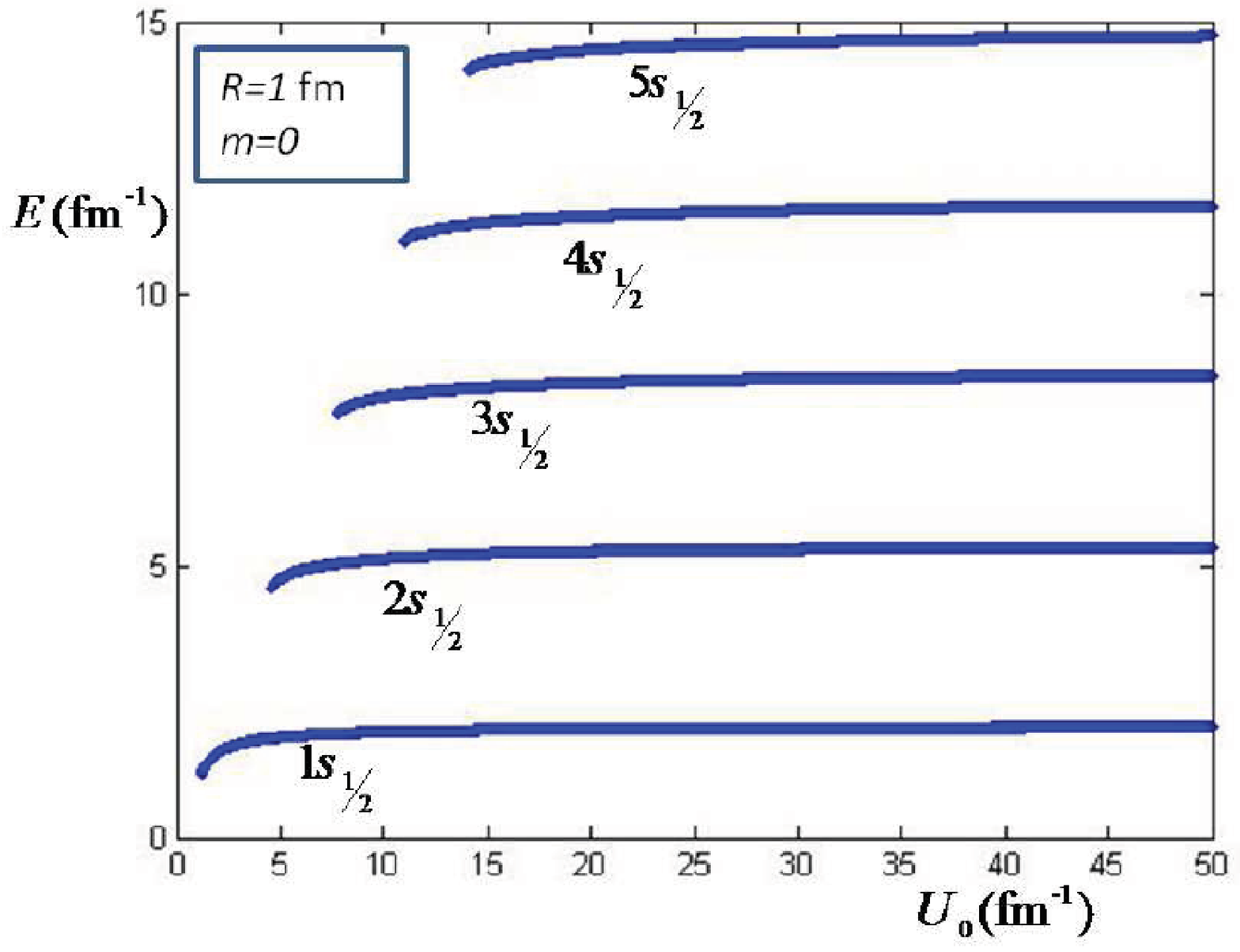}
\includegraphics[width=5.5cm]{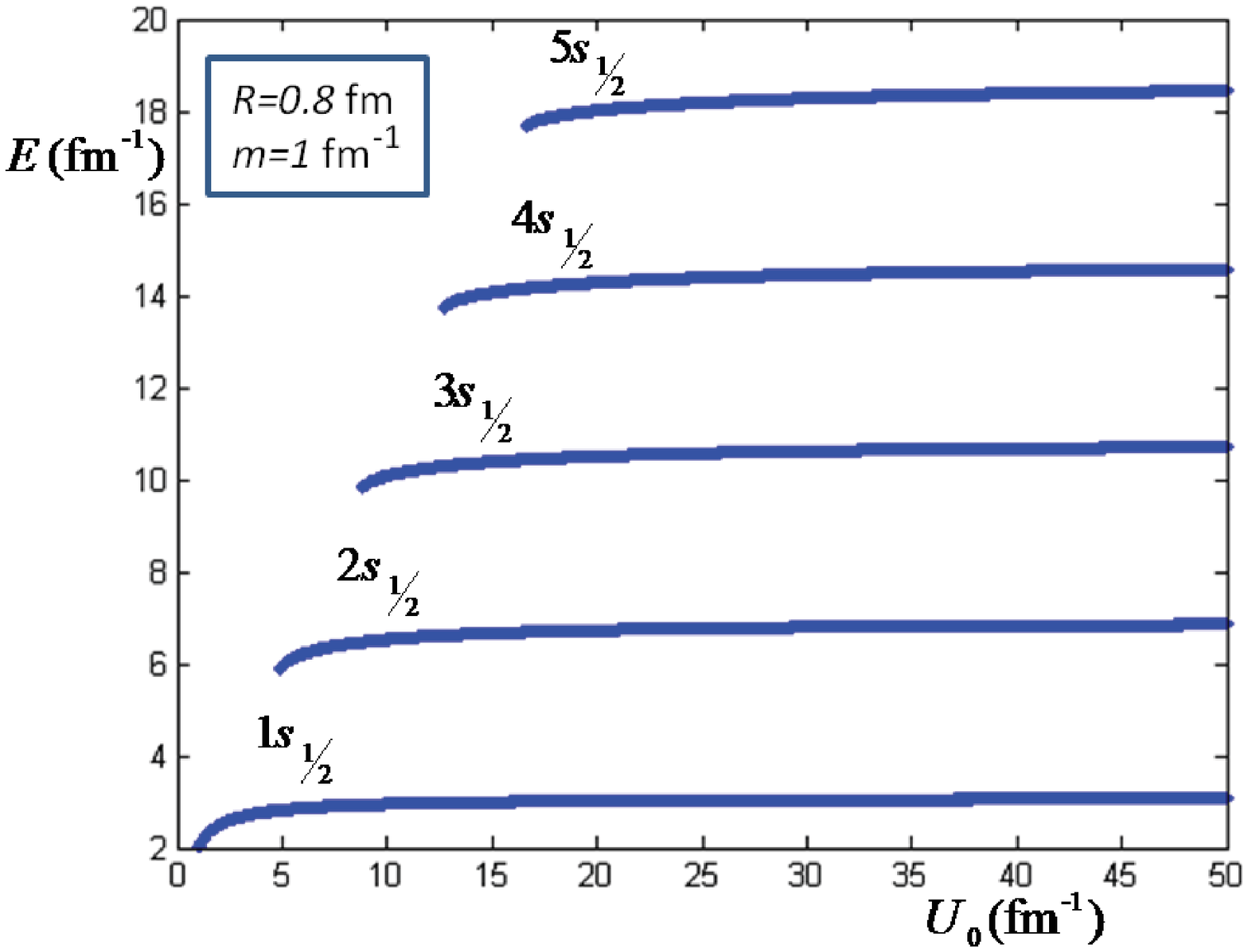}
\caption{\small
Energy levels for $s_{_{1\hspace{-.4mm}{\diagup_{\hspace{-.8mm}2}}}}$ states for a Dirac particle in a spherically symmetric potential well.}
\label{fig2}
\end{figure}

\subsection{The energy Eigenvalues for  $s_{_{1\hspace{-.4mm}{\diagup_{\hspace{-.8mm}2}}}}$ states}
The wave functions for $s_{_{1\hspace{-.4mm}{\diagup_{\hspace{-.8mm}2}}}}$ states ($\kappa=-1$) in a finite scalar potential (with no vector potential, $V_0=0$) can be written using Eqs. (\ref{8a}) and (\ref{10a}). For
two regions, inside and outside the  static spherical cavity, we have
\begin{subequations}�\label{13}
\begin{align}
\mbox{for}\quad r<R:\qquad
g_{ - 1} (r) =& Nj_0 (pr) = N\frac{{\sin (pr)}}{{pr}},\label{11a}
\\\label{11b}
f_{ - 1} (r) = & - N\frac{p}{{m + E}}j_1 (pr) =  - \frac{{Np}}{{ {m + E}}}\left[ {\frac{{\sin(pr)}}{{\left( {pr} \right)^2 }} - \frac{{\cos(pr)}}{{pr}}} \right],
\end{align}
\end{subequations}
\begin{subequations}\label{14}
\begin{align}\label{12a}
\hspace{-1.5cm}\mbox{for}\quad r>R:\qquad
g_{ - 1} (r) = &Mk_0 (qr) = M\frac{{e^{ - qr} }}{{qr}},
\\\label{12b}
f_{ - 1} (r) =& -\frac{{Mq }}{{m + E + U_0}}k_1 (qr) =  - \frac{{Mqe^{ - qr} }}{{m + E + U_0}}\left[ {\frac{1}{{qr}} + \frac{1}{{\left( {qr} \right)^2 }}} \right].
\end{align}
\end{subequations}
Using Eqs.  (\ref{11a}), (\ref{12a}) and continuity of the $g_{-1}(r)$ at $r=R$ one can find $M$ as follows:
\begin{eqnarray}
M=N\frac{q}{p}e^{qR}\sin(pR).
\end{eqnarray}
From the normalization  condition for wave functions ($\int_0 ^\infty \left[g_\kappa ^2(r) +f_\kappa ^2 (r) \right]r^2 dr= 1$) and   Eqs. (\ref{13}) and (\ref{14}), after some cumbersome calculations, we get the following expression for $N$:
\begin{eqnarray}\label{20}
&&N = \Bigg\{  \frac{R}{2p^2} + \frac{{R}}{{2(m + E)^2 }}+\frac{{\sin\left( {2pR} \right)}}{4p^3}\bigg( {\frac{{p^2 }}{{\left( {m + E} \right)^2 }} - 1} \bigg) - \frac{\sin^2 (pR)}{p^2R\left( {m + E} \right)^2 }\nonumber\\&&\hspace{2in}+ \frac{{\sin^2 (pR)}}{{2p^2 }}\left[ {\frac{1}{q} + \frac{1}{{{(m + E + U_0)^2 }}}\bigg( {q + \frac{2}{R}} \bigg)} \right] \Bigg\}^{-\frac{1}{2}}.
\end{eqnarray}
The continuity of the wave function components at $r=R$ implies
\begin{equation}\label{15}
\frac{f}{g}\left(r <R \right)\bigg|_{
 r = R}
 = \frac{f}{g}\left(r >R \right)\bigg|_{
 r = R}.
\end{equation}
Now using Eqs. (\ref{13}), (\ref{14}) and (\ref{15}), we find the following expression:
\begin{equation}\label{16}
\sqrt {\frac{{E - m}}{{E + m}}} {\cot\left( {R\sqrt {E^2  - m^2 } } \right)} - \frac{1}{{ {R\left( {E + m} \right)}}} + \sqrt {\frac{{ {m + U_0 } - E}}{{ {m + U_0  + E}}}}  + \frac{1}{{ {R\left( {m + U_0  + E} \right)}}} = 0.
\end{equation}
This eigenvalue equation gives us the energy of the particle in a scalar potential as a function of $R$ and $m$ for the states $s_{_{1\hspace{-.4mm}{\diagup_{\hspace{-.8mm}2}}}}$.

\begin{figure}[th]
\includegraphics[width=7.5cm]{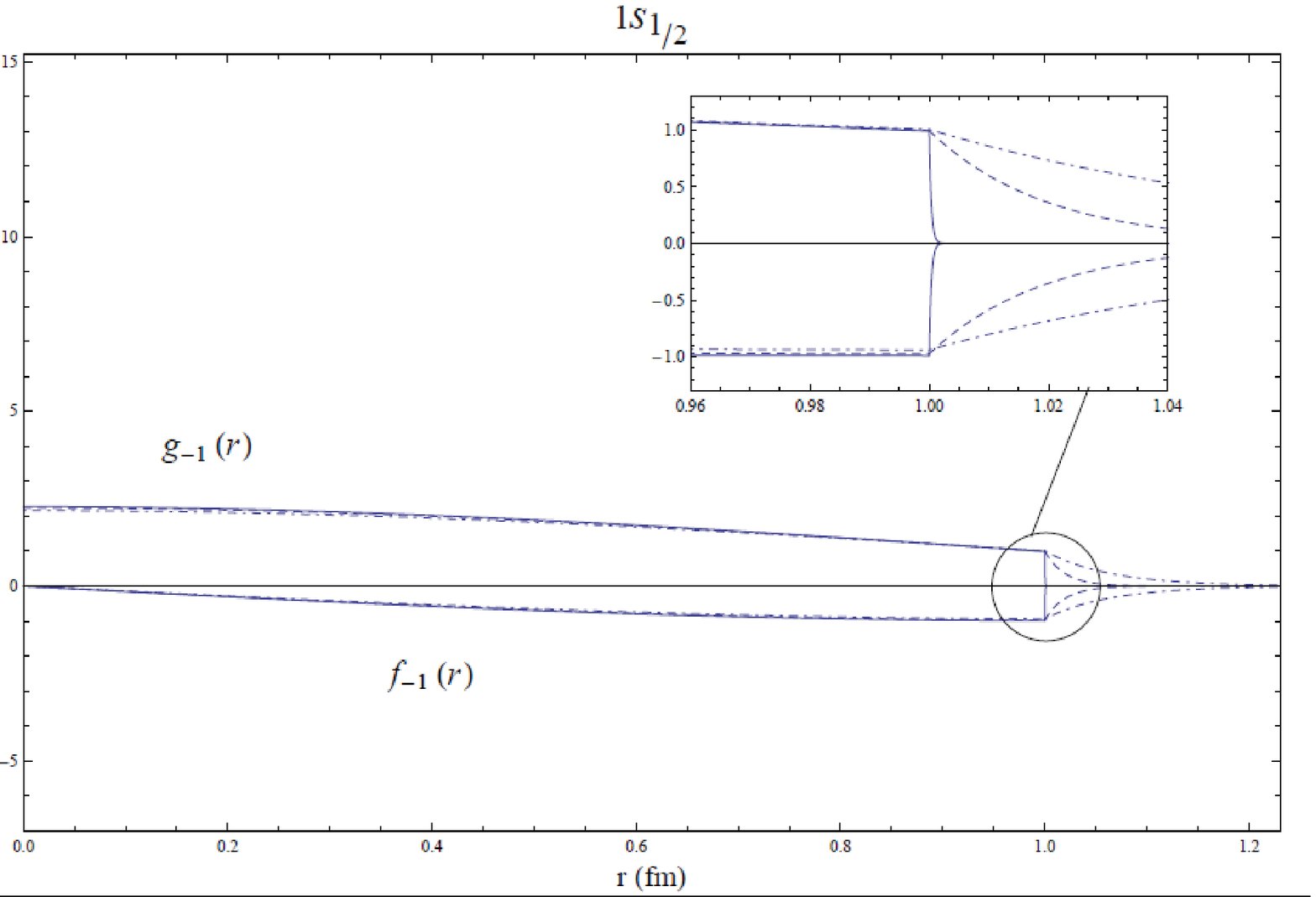}
\includegraphics[width=7.5cm]{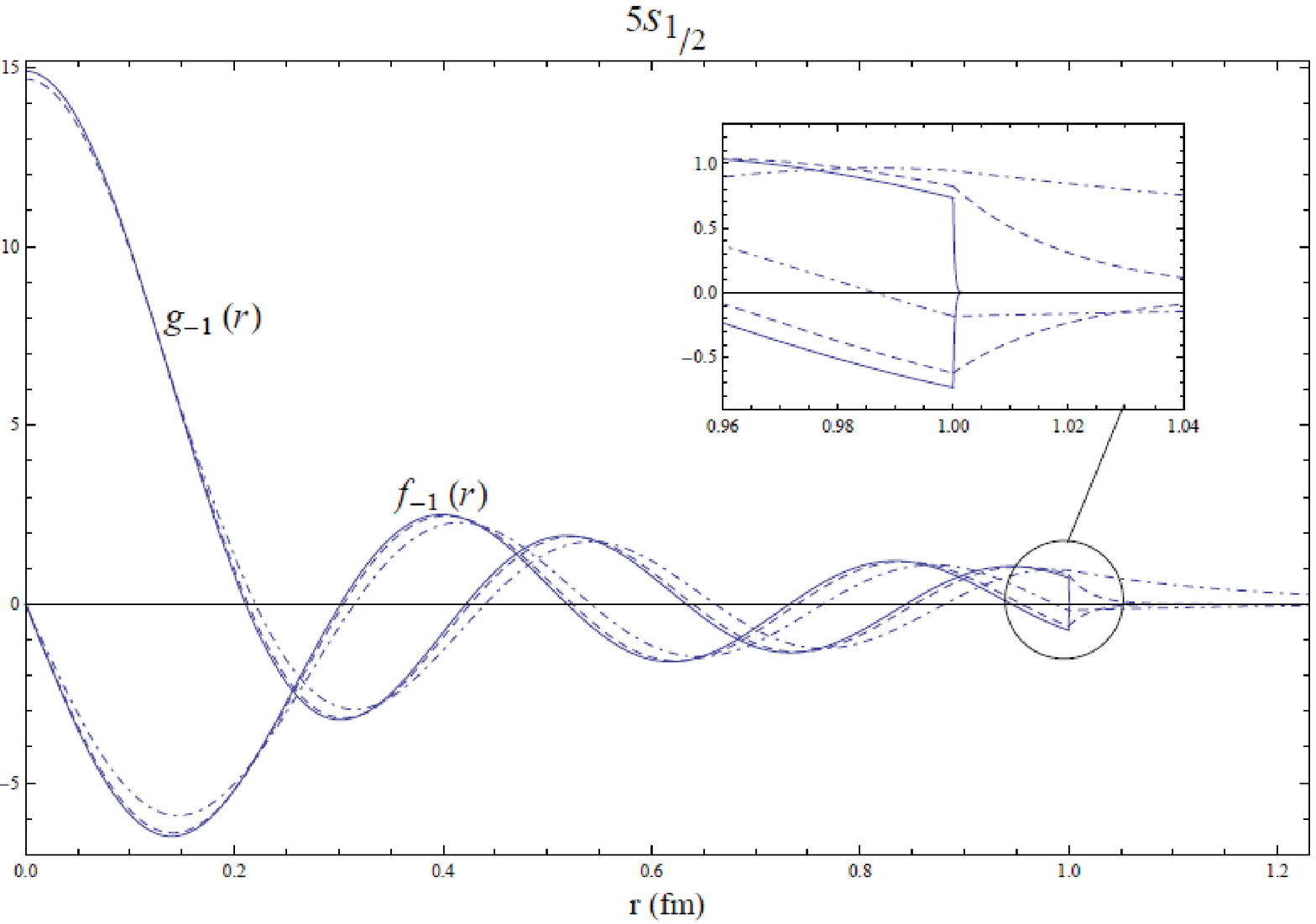}
\caption {\small
Wave functions for  the $1s_{_{1\hspace{-.4mm}{\diagup_{\hspace{-.8mm}2}}}}$ and $5s_{_{1\hspace{-.4mm}{\diagup_{\hspace{-.8mm}2}}}}$ states for $R=1$ fm and  $m=0$. Dot-dashed, dashed and solid lines denote $U_0=15$ fm$^{-1}$, $U_0=50$ fm$^{-1}$ and $U_0=\infty$, respectively. We see that as the potential well becomes deeper  (bag limit) the wave function at the boundary fall down.}
\label{fig1}
\end{figure}

We have numerically calculated the eigenvalues energy
for $s_{_{1\hspace{-.4mm}{\diagup_{\hspace{-.8mm}2}}}}$ states for different values of $R$ and $m$ and plotted them in Fig. \ref{fig2}. It is understood from numerical solutions [see Fig.  \ref{fig2} and Table \ref{table2}] that as $U_0  \to \infty$ for the case of $m=0$ and $R=1$ fm the value of the energy ground state becomes 2.0428 and for $2s_{_{1\hspace{-.4mm}{\diagup_{\hspace{-.8mm}2}}}}$ state this value is 5.3960, which are in good agreement with previously established results \cite{[5a]}(cf. Table \ref{table1}). As the potential goes to infinity the particle is completely confined inside the bag. We have also depicted the eigenvalues of the $s_{_{1\hspace{-.4mm}{\diagup_{\hspace{-.8mm}2}}}}$  states for  different values of $R$ and $m$ in Fig. \ref{fig2}, and some of the energy eigenvalues are listed in Table \ref{table2}.
In Fig. \ref{fig1} the wave functions for the $s_{_{1\hspace{-.4mm}{\diagup_{\hspace{-.8mm}2}}}}$  states for a specific case ($R=1$ fm, $m=0$) are shown. We see that, as the depth of the potential well increases, the wave function components on the  boundary of cavity become sharper, i.e. if the quarks are turned back at the edge of the nucleon by a strong interaction, the wave function will be strongly damped in that region which is in line with the MIT bag model.

\subsection{Energy eigenvalues for $p_{_{1\hspace{-.4mm}{\diagup_{\hspace{-.8mm}2}}}}$ states}

We derive the wave functions for $p_{_{1\hspace{-.4mm}{\diagup_{\hspace{-.8mm}2}}}}$ states ($\kappa=1$) using Eqs. (\ref{8b}) and (\ref{10b}), as follows:
\begin{subequations}\label{17}
\begin{align}\label{17a}
\mbox{for} \quad r<R:\quad
g_1 (r) &= N'j_1 (pr) = N'\left[ {\frac{{\sin(pr)}}{{\left( {pr} \right)^2 }} - \frac{{\cos(pr)}}{{pr}}} \right]\quad
\\\label{17b}
f_1 (r) &= \frac{{N'p}}{{{m + E}}}j_0 (pr) = \frac{{N'p}}{{{m + E} }} {\frac{{\sin\left( {pr} \right)}}{{{pr}}}},
\end{align}
\end{subequations}
\begin{subequations}\label{18}
\begin{align}\label{18a}
\mbox{for} \quad r>R:\quad
g_1 (r)& = M'k_1 (qr) = M'e^{ - qr} \left[ {\frac{1}{{ {qr} }} + \frac{1}{{\left( {qr} \right)^2 }}} \right]\quad
\\\label{18b}
f_1 (r) &=  - M'\sqrt {\frac{{ {m + U_0  - E} }}{{ {m + U_0  + E}}}} K_0 (qr) =- M'\sqrt {\frac{{ {m + U{}_0 - E} }}{{ {m + U_0  + E} }}} \frac{{e^{ - qr} }}{{qr}}.
\end{align}
\end{subequations}
As for the $s_{_{1\hspace{-.4mm}{\diagup_{\hspace{-.8mm}2}}}}$ states we can compute the normalization factor $M'$ from Eqs.  (\ref{17b})and (\ref{18b}), and imply the continuity condition of $f_1(r)$ at $r=R$, to have
\begin{eqnarray}
M'=-N'\frac{m+U_0+E}{m+E}e^{qR}\sin(pR).
\end{eqnarray}
We have also computed $N'$ from the normalization condition to be
\begin{eqnarray}
 \nonumber N' = \bigg\{ \frac{1}{{(m + E)^2 }}(\frac{R}{2} - \frac{{\sin (2pR)}}{{4p}}) - \frac{1}{{2Rp^4 }} + \frac{R}{{2p^2 }} + \frac{{\sin (2pR)}}{{4p^3 }} + \frac{{\cos (2pR)}}{{2Rp^4 }}\\
 +\frac{{\sin ^2 (pR)}}{{\mathop {(m + E)}\nolimits^2 }}[\frac{1}{{2q}} + \mathop {(m + E + U_0 )}\nolimits^2 (\frac{1}{{2q^3 }} + \frac{1}{{Rq^4 }})]\bigg\} ^{- 1/2}.
\end{eqnarray}
Now using Eqs. (\ref{15}), (\ref{17}) and (\ref{18}), we have
\begin{equation}\label{19}
 {\cot( {R\sqrt {E^2  - m^2 } } )} - \frac{{\sqrt {\left( {E - m} \right)\left( {m + U_0  + E} \right)} \left( {1 + \frac{1}{{{R\sqrt {\left( {m + U_0 } \right)^2  - E^2 } }}}} \right)}}{{\sqrt {\left( {E + m} \right)\left( {m + U_0  - E} \right)} }} - \frac{1}{{ {R\sqrt {E^2  - m^2 } } }} = 0.
\end{equation}

\begin{figure}[th]
\includegraphics[width=7.5cm]{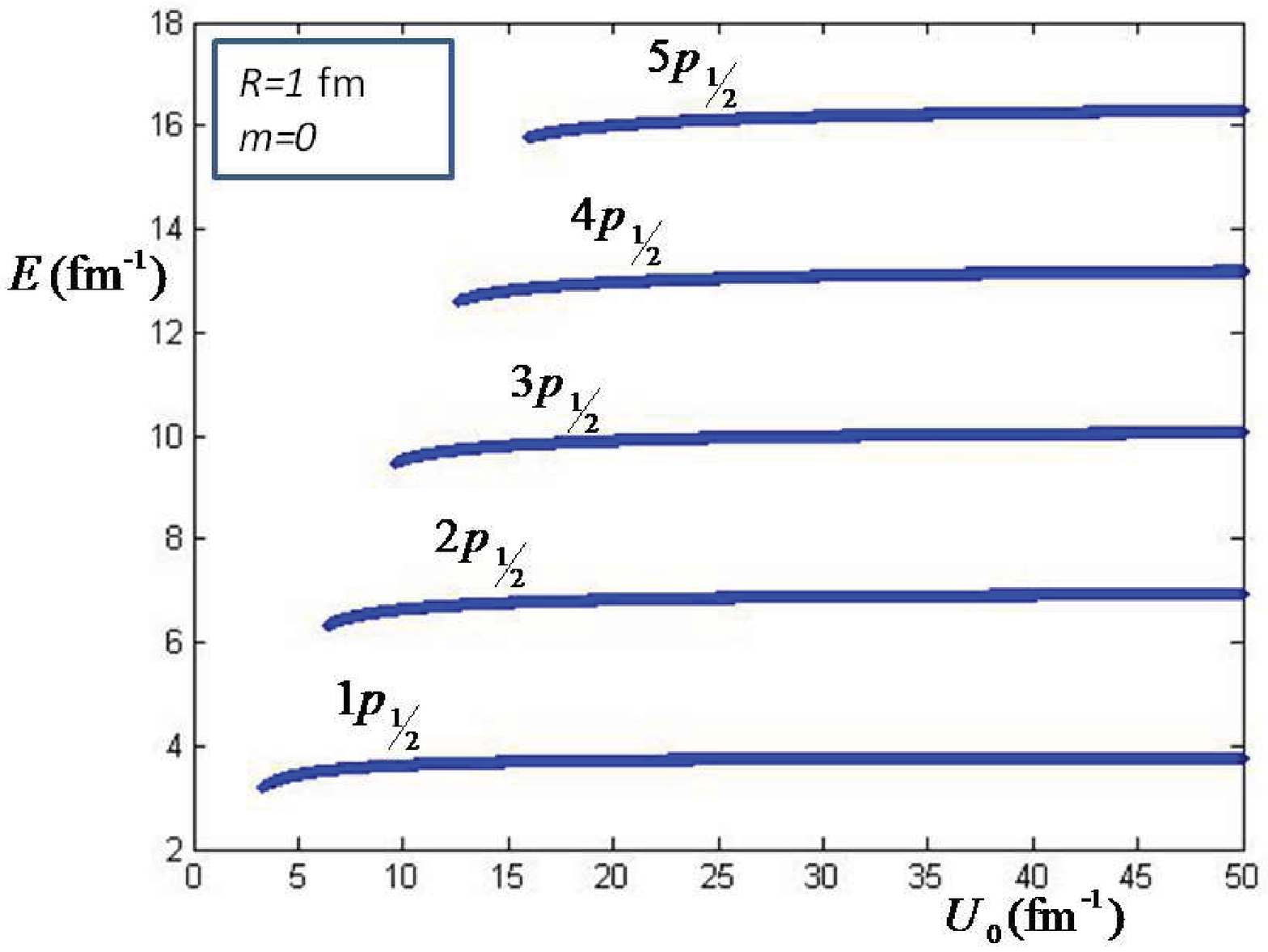}
\includegraphics[width=7.5cm]{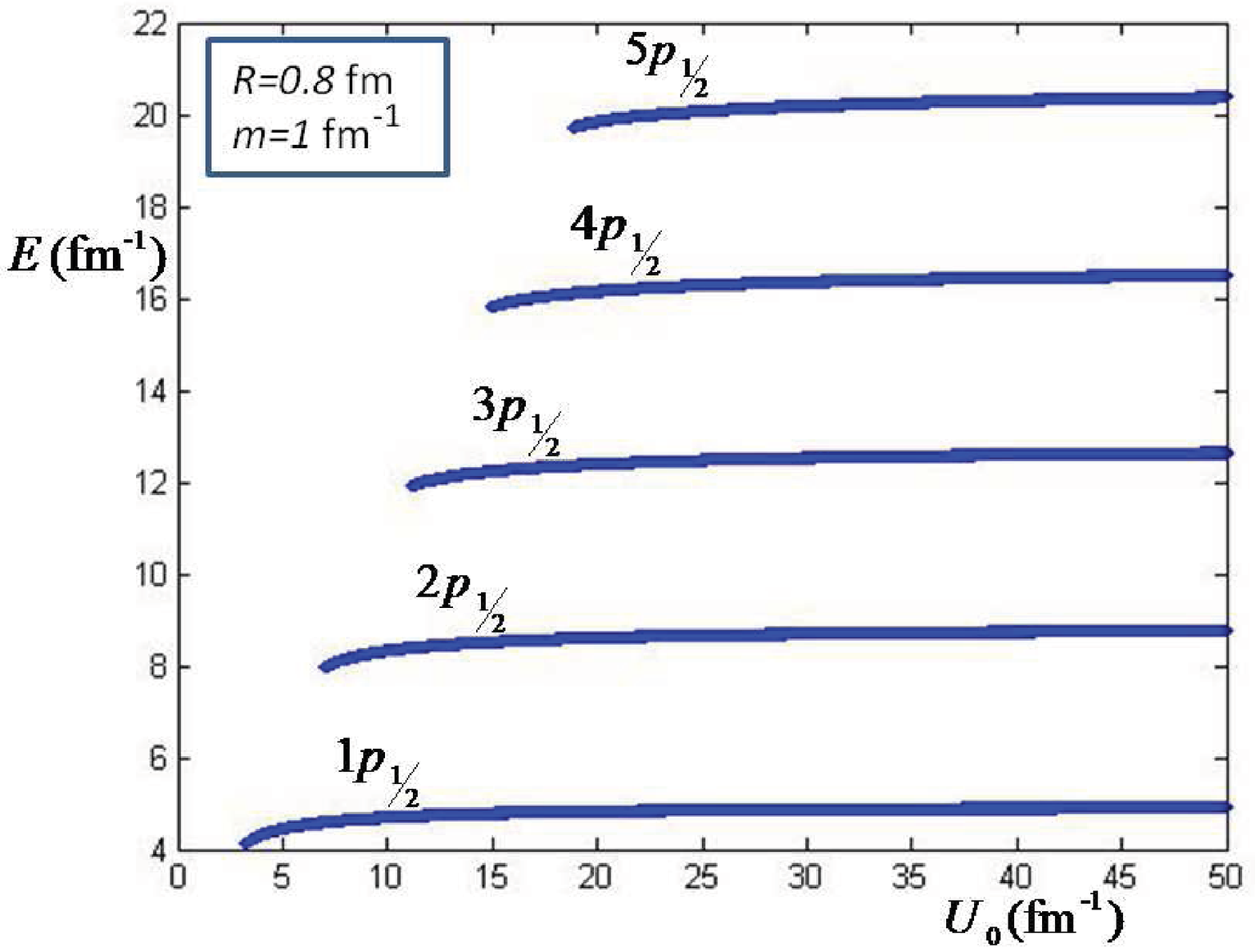}
\caption {\small
Energy levels for $p_{_{1\hspace{-.4mm}{\diagup_{\hspace{-.8mm}2}}}}$ states for a Dirac particle in a spherically symmetric potential well.}
\label{fig4}
\end{figure}

\begin{figure}[th]
\includegraphics[width=7.5cm]{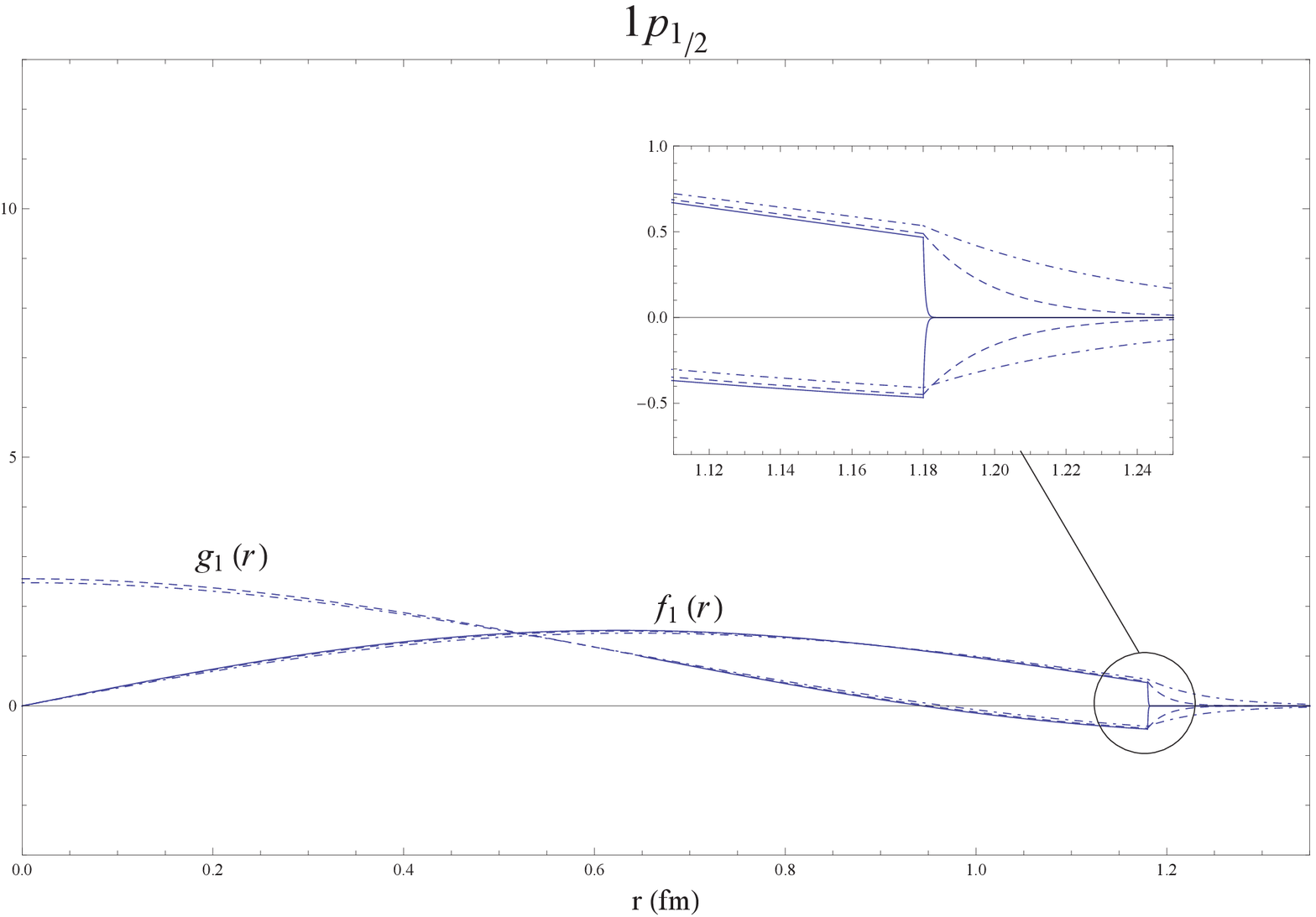}
\includegraphics[width=7.5cm]{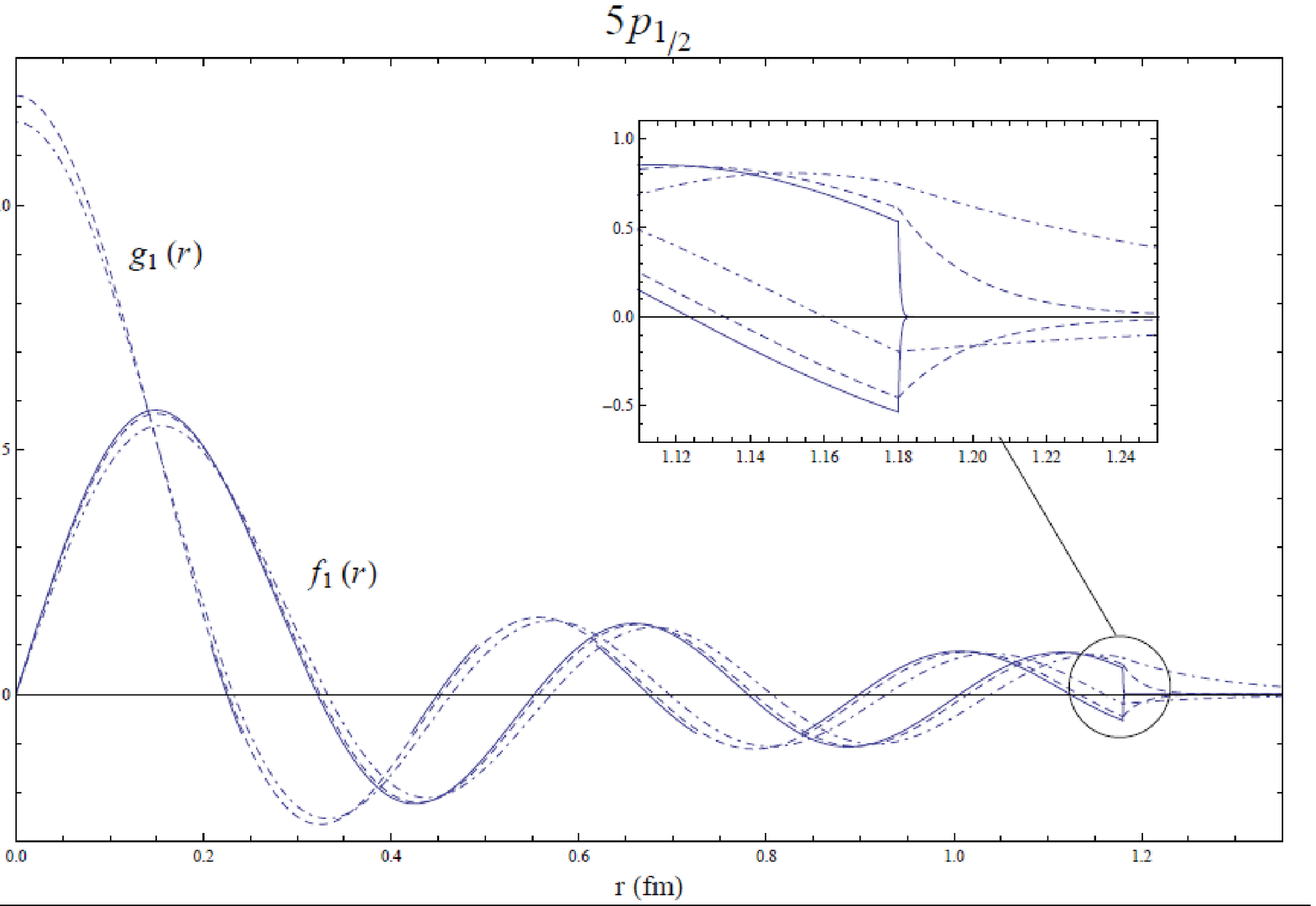}
\caption {\small
Wave functions for  the $p_{_{1\hspace{-.4mm}{\diagup_{\hspace{-.8mm}2}}}}$  and $5p_{_{1\hspace{-.4mm}{\diagup_{\hspace{-.8mm}2}}}}$states for $R=1.18$ fm and  $m=1$. Dot-dashed, dashed and solid lines denote $U_0=15$ fm$^{-1}$, $U_0=50$ fm$^{-1}$ and $U_0=\infty$, respectively. We see that as the potential well becomes deeper  (bag limit) the wave function at the boundary fall down.}
\label{fig3}
\end{figure}

This eigenvalue equation gives the particle energy for a given scalar potential as a function of $R$ and $m$ for $p_{_{1\hspace{-.4mm}{\diagup_{\hspace{-.8mm}2}}}}$ states.
We have depicted the eigenvalues for $p_{_{1\hspace{-.4mm}{\diagup_{\hspace{-.8mm}2}}}}$ states, found numerically from Eq. (\ref{19}), in Fig. \ref{fig4}. At infinite limit ($U_0  \to \infty$) we see the energy eigenvalues for $1p_{_{1\hspace{-.4mm}{\diagup_{\hspace{-.8mm}2}}}}$ and   $2p_{_{1\hspace{-.4mm}{\diagup_{\hspace{-.8mm}2}}}}$ states are 3.8115 and 7.0020 (for $m=0$ and $R=1$ fm) which are in good agreement with previously reported values (see Table \ref{table1})\cite{[5a]}. Again, in the infinite potential the particle is completely confined inside the bag. We have listed some of the energy eigenvalues in Table \ref{table3}.
We have also illustrated the wave functions of  the $p_{_{1\hspace{-.4mm}{\diagup_{\hspace{-.8mm}2}}}}$  states for $R=1.18$ fm and  $m=1$ fm$^{-1}$ in Fig. \ref{fig3}.

Comparing the results of energy eigenvalues for the quark particle in the different static spherical cavity radius $R$, and different quark masses $m$, [Figs. \ref{fig2} and \ref{fig4}, and Tables \ref{table2} and \ref{table3}] one can see that an increase in $R$ decreases the energy eigenvalues. Also, an increase in $m$ leads to the increase in energy. Therefore, the energy eigenvalues of a quark particle confined in a static spherical cavity highly depend on its mass and the radius of the cavity. These results are in agreement with the relativistic statistical mechanics. The quark particles inside the bag behave similar to a relativistic gas, so that the quarks kinetic pressure is equal to the pressure of the gas \cite{[10],path}.

\subsection{Additional remarks about the sharpness of the wave function}

In the case of the scalar potential discussed in previous sections, it is also interesting to investigate the wave function continuity at $r=R$ for  any $\kappa$. To study the situation, we use Eqs. (\ref{7}) and (\ref{7.1})  along with the definitions $G_\kappa  \left( r \right) = rg_\kappa  \left( r \right)$ and $F_\kappa  \left( r \right) = rf_\kappa  \left( r \right)$ for $r>R$, then we have
\begin{equation}\label{21}
\frac{{d^2 G_\kappa  (r)}}{{dr^2 }} - \frac{{\kappa \left( {\kappa  + 1} \right)}}{{r^2 }}G_\kappa  (r) - \left[ {\left( {m + U(r)} \right)^2  - E^2 } \right]G_\kappa  (r) - \frac{{{\frac{{dU(r)}}{{dr}}\frac{{dG_\kappa  (r)}}{{dr}}} }}{{ {m + E + U(r)} }} - \frac{{\kappa {G_\kappa  (r)\frac{{dU(r)}}{{dr}}}}}{{r\left[ {m + E + U(r)} \right]}} = 0.
\end{equation}

Wherever $U(r)$ has a sharp point,  $\frac{{dU\left( r \right)}}{{dr}}$ has a certain jump. As $U(r)$  goes to infinity we have the Dirac $\delta$ potential, whose first derivative is a larger infinity. In this case, to compensate for such a large jump, $\frac{{d^2 G_\kappa  \left( r \right)}}{{dr^2 }}$  should have the same large jump. Therefore, we can conclude that $\frac{{d G_\kappa  \left( r \right)}}{{dr }}$
has a jump.

We know that $G_\kappa  \left( r \right)$
 is a continuous wave function. Now integrating Eq. (\ref{21}) in the small interval  $\left[ {R - \varepsilon ,R + \varepsilon } \right]$, and taking the limit $\varepsilon  \to  \circ $ lead to the zero contribution for the second and third terms. On the other hand, we have
\begin{equation}\label{22}
\int_a^b {\delta (r - R)}F(r)dr = \left\{ {\begin{array}{cc}
   F(R)&\qquad r \in \left[ {a,b} \right]  \\
   \frac{{F(R)}}{2}&\qquad r =a \;\mbox{or} \;b \\
   0 &\qquad r \notin \left[a,b\right],
\end{array}} \right.
\end{equation}
and
\begin{equation}\label{23}
\frac{{dU\left( r \right)}}{{dr}} = U_0  \delta \left( {r - R} \right).
\end{equation}
Using Eqs. (\ref{22}) and (\ref{23}) we can compute the remaining terms of integration to get
\begin{equation}\label{24}
\mathop {\lim }\limits_{\varepsilon  \to 0} \left( {\frac{{dG_\kappa  (r)}}{{dr}}\bigg|_{R - \varepsilon }^{R + \varepsilon } } \right) = \frac{{U_0 }}{{ {m + E + \frac{{U_0 }}{2}}}}{\overline {G' _\kappa}}  (R) + \frac{{\kappa U_0 }}{{R\left[ {m + E + \frac{{U_0 }}{2}} \right]}}G_\kappa  (R),
\end{equation}
where $\overline {G'} _\kappa  (R)$ denotes the mean value of the first derivative of $G_\kappa  (r)$ on the wall. This relation gives the sharpness of the wave function $G_\kappa  (r)$ while crossing the wall.
In a similar way, one can obtain the sharpness of $F_\kappa(r)$
\begin{equation}\label{25}
 \lim \limits_{\varepsilon  \to 0} \left(\frac{{dF_\kappa( r)}}{{dr}} \bigg|^{R + \varepsilon }_
   {R - \varepsilon } \right) = \frac{{U_0 }}{{ {m  - E + \frac{{U_ \circ  }}{2}}}} \overline  {F'_\kappa} ( R )- \frac{{\kappa U_0  }}{{R\left[ {m - E + \frac{{U_ \circ  }}{2}} \right]}}F_\kappa (R).
\end{equation}

\section{Conclusions}\label{sec4}
In this paper we have computed the energy eigenvalues for a Dirac particle in a scalar potential with full spherical symmetry.  Although in finite potential this model is not compatible with quark confinement, however here we observed the evolution of the wave functions as $U_0\to\infty$, where we recovered the MIT bag model results. We have found the components of the wave function and depicted them in Figs. \ref{fig1} and \ref{fig3}. As a conclusion energy eigenvalues of a particle for any potential value are positive values. For both cases of  $s_{_{1\hspace{-.4mm}{\diagup_{\hspace{-.8mm}2}}}}$ and $p_{_{1\hspace{-.4mm}{\diagup_{\hspace{-.8mm}2}}}}$ states in infinite potential the particle is completely confined inside the bag. It is pleasing that all of our results are in good agreement with that which exists in the literature for the infinite potential limit (the MIT bag model). We also have additional remarks about the continuity and sharpness of the wave function. In the case of the Dirac equation with a finite  potential, the  wave function is continuous, however it has a sharp point at $r=R$. We have calculated the sharpness of the wave function for any $\kappa$, and we see as the potential goes to the infinity ($U_0\to\infty$), the wave function becomes discontinuous. We saw the results preserved the relativistic statistical mechanics, so that the quark particle inside the bag behaves as a relativistic gas.

\vskip20pt\noindent {\large {\bf Acknowledgement}}\vskip5pt\noindent
The authors are extremely grateful to S.S. Gousheh for all his ultra-prompt help, which
greatly improved this work.

\newpage

\begin{table}
\begin{tabular}{|c|c|c|c|c|c|c|c|}\hline
	States&Potential &$m$=0&$m$=1\ fm$^{-1}$&$m$=0&$m$=1\ fm$^{-1}$&$m$=0&$m$=1 \ fm$^{-1}$\\
&$U_0$\ (fm$^{-1})$&$R$=0.8\ fm&$R$=0.8\ fm&$R$=1\ fm&$R$=1\ fm&$R$=1.18\ fm&$R$=1.18\ fm\\
\hline
\hline
&$15$&2.4492&2.9949&1.9758&2.5342&1.6829&2.2529	\\
$1s_{_{1\hspace{-.4mm}{\diagup_{\hspace{-.8mm}2}}}}$&$50$
&2.5218&3.0625&2.0225&2.5772&1.7166&2.2835\\
&$\infty$&2.5535&3.0932&2.0428&2.5966&1.7312&2.2973\\
\hline
&$15$&6.4655&6.6608&5.2178&5.4262&4.4451&4.6663\\$2s_{_{1\hspace{-.4mm}{\diagup_{\hspace{-.8mm}2}}}}$&$50$
&6.6614&6.8464&5.3424&5.5437&4.5344&4.7501\\
&$\infty$&6.7450&6.9284&5.3960&5.5960&4.5729&4.7875\\
\hline
&$15$&10.2508&10.3833&8.2845&8.4197&7.0611&7.2029\\$3s_{_{1\hspace{-.4mm}{\diagup_{\hspace{-.8mm}2}}}}$&$50$
&10.5885&10.7020&8.4922&8.6164&7.2078&7.3417\\
&$\infty$&10.7219&10.8336&8.5776&8.7004&7.2691&7.4019\\
\hline
&$15$&13.9259&14.0606&11.3090&11.4184&9.6498&9.7590\\$4s_{_{1\hspace{-.4mm}{\diagup_{\hspace{-.8mm}2}}}}$&$50$
&14.4868&14.5694&11.6192&11.7094&9.8620&9.9594\\
&$\infty$&14.6706&14.7509&11.7365&11.8251&9.9462&10.0423\\	
\hline
&$15$&---    &---&14.2597&14.3819&12.2134&12.3104\\$5s_{_{1\hspace{-.4mm}{\diagup_{\hspace{-.8mm}2}}}}$&$50$
&18.3745&18.4400&14.7382&14.8094&12.5096&12.5863\\
&$\infty$&18.6098&18.6725&14.8878&14.9572&12.6168&12.6921\\
\hline
\end{tabular}
\caption {\small
Energy eigenvalues (in fm$^{-1}$) for the $s_{_{1\hspace{-.4mm}{\diagup_{\hspace{-.8mm}2}}}}$ states for a Dirac particle in a spherically symmetric potential well.}
\label{table2}
\end{table}

\begin{table}
\begin{tabular}{|c|c|c|c|c|c|c|c|}\hline
	States&Potential &$m$=0&$m$=1\ fm$^{-1}$&$m$=0&$m$=1\ fm$^{-1}$&$m$=0&$m$=1 \ fm$^{-1}$\\
&$U_0$\ (fm$^{-1})$&$R$=0.8\ fm&$R$=0.8\ fm&$R$=1\ fm&$R$=1\ fm&$R$=1.18\ fm&$R$=1.18\ fm\\
\hline
\hline
&$15$& 4.5719&  4.7840  &3.6878   & 3.9207  & 3.1410& 3.3928\\
$1p_{_{1\hspace{-.4mm}{\diagup_{\hspace{-.8mm}2}}}}$&$50$
& 4.7056& 4.9098 &  3.7738  & 4.0007  &  3.2030& 3.4498\\
&$\infty$& 4.7644&4.9669 &3.8115 &4.0370  & 3.2301 &3.4756\\
\hline
&$15$&  8.3835 & 8.5171  & 6.7686 & 6.9107 &   5.7671&5.9189\\$2p_{_{1\hspace{-.4mm}{\diagup_{\hspace{-.8mm}2}}}}$&$50$
& 8.6440 &  8.7642 & 6.9325  &  7.0659 &  5.8840 &6.0293\\
&$\infty$&8.7525 &8.8709   & 7.0020  &  7.1341  & 5.9339  &6.0781\\
\hline
&$15$& 12.1192 &   12.2330&  9.8080    &  9.9169 &8.3629&8.4755 \\$3p_{_{1\hspace{-.4mm}{\diagup_{\hspace{-.8mm}2}}}}$&$50$
&12.5457 &  12.6315 & 10.0621  &  10.1569  &  8.5403 &8.6434\\
&$\infty$& 12.7042 &  12.7879 &  10.1633 & 10.2567  &  8.6130 &8.7149\\
\hline
&$15$& ---&---&  12.8059 &  12.9075 & 10.9389  & 11.0341  \\$4p_{_{1\hspace{-.4mm}{\diagup_{\hspace{-.8mm}2}}}}$&$50$
& 16.4353 &  16.5026 &  13.1823 &  13.2561 & 11.1888  &11.2689\\
&$\infty$&16.6445  &  16.7093&13.3156 & 13.3878  & 11.2844  & 11.3632\\	
\hline
&$15$&---& --- &  --- &  --- &  13.4807 &  13.5764\\$5p_{_{1\hspace{-.4mm}{\diagup_{\hspace{-.8mm}2}}}}$&$50$
& 20.3185 & 20.3744  & 16.2979  &  16.3588 & 13.8337  &13.8995\\
&$\infty$& 20.5799 & 20.6327  & 16.4639  &  16.5227 & 13.9525  &14.0167 \\
\hline
\end{tabular}
\caption {\small
Energy eigenvalues (in fm$^{-1}$)for the $p_{_{1\hspace{-.4mm}{\diagup_{\hspace{-.8mm}2}}}}$ states for a Dirac particle in a spherically symmetric potential well.}
\label{table3}
\end{table}

\end{document}